\RequirePackage{amsmath}
\documentclass[runningheads]{llncs}

\usepackage{ecir25-set-encoder-frame}
\graphicspath{{./ecir25-set-encoder-figures/}}

\begin{document}

\title{Set-Encoder: Permutation-Invariant\texorpdfstring{\\}{} Inter-Passage Attention for Listwise Passage Re-Ranking with Cross-Encoders}
\titlerunning{Set-Encoder: Permutation-Invariant Inter-Passage Attention}

\author{
  Ferdinand Schlatt\inst{1}\orcidID{0000-0002-6032-909X}
  \and Maik Fr{\"o}be\inst{1}\orcidID{0000-0002-1003-981X}
  \and Harrisen Scells\inst{2,6}\orcidID{0000-0001-9578-7157}
  \and \\
  Shengyao Zhuang\inst{3,4}\orcidID{0000-0002-6711-0955}
  \and Bevan Koopman\inst{3}\orcidID{0000-0001-5577-3391}
  \and Guido Zuccon\inst{4}\orcidID{0000-0003-0271-5563}
  \and \\
  Benno Stein\inst{5}\orcidID{0000-0001-9033-2217}
  \and Martin Potthast\inst{2,6,7}\orcidID{0000-0003-2451-0665}
  \and Matthias Hagen\inst{1}\orcidID{0000-0002-9733-2890}
}
\authorrunning{Schlatt et al.}
\institute{
Friedrich-Schiller-Universit{\"a}t Jena
\and University of Kassel
\and CSIRO \\
\and University of Queensland
\and Bauhaus-Universit{\"a}t Weimar \\
\and hessian.AI
\and ScadDS.AI
\\\email{ferdinand.schlatt@uni-jena.de}
}

\maketitle
\setcounter{footnote}{0}

\begin{abstract}
Existing cross-encoder models can be categorized as pointwise, pairwise, or listwise. Pairwise and listwise models allow passage interactions, which typically makes them more effective than pointwise models but less efficient and less robust to input passage order permutations. To enable efficient permutation-invariant passage interactions during re-ranking, we propose a new cross-encoder architecture with inter-passage attention: the Set-Encoder. In experiments on TREC Deep Learning and TIREx, the Set-Encoder is as effective as state-of-the-art listwise models while being more efficient and invariant to input passage order permutations. Compared to pointwise models, the Set-Encoder is particularly more effective when considering inter-passage information, such as novelty, and retains its advantageous properties compared to other listwise models. Our code is publicly available at \url{https://github.com/webis-de/ECIR-25}.
\end{abstract}

\section{Introduction}

Existing cross-encoders for passage re-ranking~\cite{nogueira:2019,nogueira:2020a,nogueira:2020b,pradeep:2021,pradeep:2022,pradeep:2023a,pradeep:2023,sun:2023,tamber:2023,zhuang:2022} lack a central desirable property of learning-to-rank models~\cite{pang:2020}: permutation-invariant passage interactions. Passage interactions help to improve effectiveness through information exchange, while permutation invariance ensures that the same ranking is output independent of the ordering of the input passages.

Pointwise cross-encoders process passages independently of each other~\cite{nogueira:2020a,nogueira:2020b,pradeep:2022,zhuang:2022} and thus are permutation-in\-variant by design but cannot model passage interactions. Pairwise~\cite{nogueira:2019,pradeep:2021} and listwise cross-encoders~\cite{sun:2023,pradeep:2023a,pradeep:2023,tamber:2023} enable passage interactions by concatenating the passages as a single input sequence. This way, pair- or listwise re-rankers are often more effective than pointwise ones but the derived rankings depend on the concatenation order. To avoid inconsistent rankings for different input orderings, many (or all) permutations are re-ranked and fused to a final ranking; visualized in Figure~\ref{cross-encoder-architectures} (bottom).

To directly model passage interactions in a permutation-invariant way, we propose a new cross-encoder architecture: the \emph{Set-Encoder}; visualized in Figure~\ref{cross-encoder-architectures} (top). Instead of concatenating the input passages, the Set-Encoder processes them like a pointwise cross-encoder as different sequences in a batch. Passage interactions are enabled by letting each sequence aggregate information in special [INT]~interaction embedding tokens that all other sequences can attend to. The [INT]~tokens can be interpreted as lightweight bridges which share information between passages. To ensure permutation-invariance, the Set-Encoder assigns every [INT]~token the same positional encoding. We call this new attention pattern \emph{inter-passage attention}.

\begin{figure}[t]
\centering
\includegraphics{cross-encoder-architectures-landscape.ai}
\caption{Comparison of our Set-Encoder architecture~(top) with state-of-the-art listwise re-rankers~(bottom) for three input passages. List-wise re-rankers concatenate the input passages, leading to potentially inconsistent rankings for different concatenation orderings. Many (or all) permutations are thus re-ranked for optimization. Instead, the Set-Encoder uses novel [INT] tokens for permutation-invariant inter-passage attention.}
\label{cross-encoder-architectures}
\end{figure}

In experiments on the TREC~2019 and~2020 Deep Learning tracks~\cite{craswell:2019,craswell:2020} and the TIREx framework~\cite{frobe:2023}, the Set-Encoder is as effective as state-of-the-art re-rankers~\cite{zhuang:2022,pradeep:2023,sun:2023}. At the same time, the Set-Encoder is robust to input permutations and orders of magnitude more efficient. Interestingly, contrasting previous work~\cite{nogueira:2019,nogueira:2020b,pradeep:2021,tamber:2023,zhuang:2022,pradeep:2023}, we find listwise interactions to not really be necessary in the Deep Learning and TIREx scenarios. A pointwise model fine-tuned with the currently best available data~\cite{schlatt:2024} is as effective as state-of-the-art listwise models and the Set-Encoder. To still demonstrate the potential advantages of listwise models, we fine-tune the Set-Encoder to re-rank passages according to both relevance and novelty and find it to then be more effective than its pointwise counterpart and other state-of-the-art listwise models.

\section{Related Work}
\label{related-work}

Pre-trained transformer-based models can roughly be divided into two types~\cite{lin:2022}: bi-encoders and cross-encoders. Bi-encoders embed queries and passages independently so that pre-computed passage representations can be indexed. At query time, only the query needs to be embedded and can then efficiently be compared with the indexed passage embeddings~\cite{reimers:2019}. Cross-encoders instead embed queries and passages together, allowing for interactions~\cite{nogueira:2020a}---for brevity, we refer to all such transformer-based models as cross-encoders, be they encoder-only~\cite{nogueira:2020a}, decoder-only~\cite{zhuang:2022,pradeep:2023a,pradeep:2023}, or encoder--decoders~\cite{nogueira:2020b,zhuang:2022}.

This direct query--passage interaction makes cross-encoders particularly effective. However, current pairwise and listwise cross-encoders lack a central property that previously improved feature-based learning-to-rank models~\cite{lee:2019a,pang:2020,pobrotyn:2020,pasumarthi:2020,buyl:2023}: input permutation-invariant interactions. The rank order of current list and pairwise cross-encoders is sensitive to the input passage ordering and often fuse multiple rankings for different input permutations to be effective.

For example, duo cross-encoders embed a query together with two passages to predict which passage of the pair should be ranked higher~\cite{pradeep:2021,nogueira:2019}. But the predictions of duo cross-encoders are neither symmetric nor transitive~\cite{gienapp:2022}, so that they must process all pairs and twice for maximum effectiveness. More recent decoder-only LLMs that simultaneously re-rank up to 20~passages~\cite{sun:2023,pradeep:2023a,pradeep:2023} are also sensitive to the input ordering, so that they usually fuse the re-rankings for several input permutations for maximum effectiveness~\cite{tang:2023}.

Existing cross-encoders that model passage interactions are not permutation-invariant because they concatenate passages into a single input sequence. Our Set-Encoder follows a different strategy and processes passages in parallel. To enable passage interactions while ensuring permutation invariance, we add a new dedicated interaction token ([INT]) to each input sequence. The Set-Encoder allows all passages to attend to the interaction tokens of all other passages. This new inter-passage attention pattern is somewhat similar to the attention pattern of the Fusion-in-Encoder model proposed for question answering~\cite{kedia:2022}. The Fusion-in-Encoder model enables passage interactions within an encoder by adding additional so-called global tokens to which all passage tokens can attend to, and the global tokens can attend to all passage tokens. However, no direct interaction between passage tokens is possible. In contrast, our Set-Encoder allows for direct passage interactions as all passage tokens can attend to the interaction tokens of every other passage.

\section{The Set-Encoder Model}
\label{set-encoder}

Our Set-Encoder introduces inter-passage attention, a novel attention pattern to model permutation-invariant interactions between passages. Inter-passage attention addresses three challenges:
\Ni
input passages must be able to attend to each other,
\Nii
the interactions between the passages must not encode any positional information about the passage ordering, and
\Niii
the interactions must be ``lightweight'' enough for efficient fine-tuning.

Previous work has addressed the first challenge (attention between passages) by concatenating the query and multiple passages into one input sequence, visualized in Figure~\ref{cross-encoder-architectures} (bottom). However, concatenation violates the second challenge: the language model's positional encodings (used to determine a token's position) span across passage boundaries and thus encode the order of passages in the sequence. Furthermore, concatenation also violates the third challenge (efficiency), as the language model's computational cost scales quadratically with the length of the input sequence.

Instead of concatenating passages, our new inter-passage attention pattern processes the passages in parallel with individual input sequences per passage, as visualized in Figure~\ref{cross-encoder-architectures} (top). Each input sequence's positional encodings start from zero, so no information about the passage order is encoded (second challenge). To still let the passages attend to each other (first challenge), we use a special passage interaction [INT]~token that aggregates semantic information about its passage and to which tokens from other sequences can attend. The Set-Encoder allows passage interactions solely through these [INT]~tokens, making inter-passage attention computationally efficient (third challenge).

\subsection{Permutation-Invariant Input Encoding}

Standard pointwise cross-encoders compute a relevance score for query--passage pairs~$(q, d)$ given as token sequences~$t_{1_q} \ldots t_{m_q}$ and~$t_{1_d} \ldots t_{m_d}$. In this work we use BERT-style encoding~\cite{devlin:2019}, meaning the final tokens $t_{m_q}$~and~$t_{m_d}$ are special [SEP]~separator tokens. The concatenated sequence is prepended by a special [CLS]~classification token $t_c$. The resulting input sequence $t_c \ t_{1_q} \ldots t_{m_q} \ t_{1_d} \ldots t_{m_d}$ is then passed through a transformer-based encoder model. Finally, the relevance score of~$d$ for~$q$ is computed by a linear transformation on the final contextualized embedding of the [CLS]~token~$t_c$.

The Set-Encoder instead receives a query~$q$ and a set of passages $\{d_1, \ldots, d_k\}$ as input and computes a relevance score for each passage. To this end, the Set-Encoder builds a set of input sequences by prepending a [CLS]~token, an [INT]~token, and the query sequence to each passage individually. The set of input sequences is then processed simultaneously, similar to batched processing with a standard cross-encoder, but with attention from an input sequence's tokens to the [INT]~tokens of the other input sequences (see Section~\ref{inter-passage-attention}). The batched input sequences are passed through an encoder, and the relevance scores of the passages~$d_i$ for~$q$ are computed by applying a linear transformation to the passage's final [CLS]~token embedding.

Our batched input encoding is permutation-invariant because each input sequence's positional encodings start from zero. For example, the [INT]~token is always at position $1$ and the first token of each passage is at position $m_q + 1$. The model then cannot distinguish between different permutations of the input sequences and, therefore, also not between different permutations of the passages.

\subsection{Inter-Passage Attention} \label{inter-passage-attention}

The Set-Encoder allows every sequence to attend to the [INT]~tokens of every other sequence. Our intuition is that these [INT]~tokens aggregate the semantic information from their sequence and can share it with all other sequences.

This hypothesis is motivated by the token-based semantic information aggregation in many model architectures. For instance, in bi-encoders, the [CLS]~token captures a query or passage's semantic information, and in a standard pointwise cross-encoder, the [CLS]~token aggregates query--passage information to compute a relevance score. We hypothesize that our additional [INT]~token can also aggregate semantic information and share this information with other passages. We also tested directly using the [CLS]~token for passage interactions and found this variant of passage interaction to produce effective re-rankers but being incapable of learning passage interactions (see Section~\ref{sec:novelty-ranking}). To implement inter-passage attention, we modify the original attention function of the transformer~\cite{vaswani:2017}:
\[
  \text{Attention}(Q, K, V) = \text{softmax} \left( \dfrac{QK^T}{\sqrt{h}} \right) V\,,
\]
where $Q$, $K$, and $V$ are matrices of embedding vectors, in which each column vector corresponds to a token from the encoder's input sequence.

The inter-passage attention's input has vector matrices $Q^{(i)}$, $K^{(i)}$, and~$V^{(i)}$ for each passage~$d_i$'s input sequence. To allow the $i$-th input sequence to attend to the [INT]~tokens of all other input sequences, we append the [INT]~token's embedding vectors from the other input sequences' embedding matrices~$K^{(j)}$ and~$V^{(j)}$ to the input sequence's embedding matrices~$K^{(i)}$ and~$V^{(i)}$.

Specifically, let $\bar{K}^{(i)} = [K^{(j)}_2 : j \neq i]$ and $\bar{V}^{(i)} = [V^{(j)}_2 : j \neq i]$ be the matrices concatenating the [INT]~token's embedding vectors from~$K^{(j)}$ and~$V^{(j)}$ of every passage $d_j \neq d_i$, with $[\cdot\,\cdot]$ denoting column-wise vector\,/\,matrix concatenation (i.e., $[M M']$ is equal to a matrix whose ``left'' columns come from~$M$ and whose ``right'' columns come from~$M'$). Appending the [INT]~tokens from the other passages' embedding vectors to the sequence matrices, we obtain the new inter-passage attention pattern as:  $\text{Attention}(Q^{(i)}, [K^{(i)} \bar{K}^{(i)}], [V^{(i)} \bar{V}^{(i)}])$.

\subsection{Fine-tuning}

We follow \citet{schlatt:2024} to fine-tune the Set-Encoder in two stages. In the first stage, the model is fine-tuned on MS~MARCO~\cite{nguyen:2016} relevance labels using InfoNCE loss~\cite{oord:2019}:
\[
  \mathcal{L}_{\text{InfoNCE}} = -\log \frac{\exp (s_+)}{\sum_{i=1}^{k} \exp (s_i)} \,.
\]
For a single query~$q$, the loss is computed over the predicted relevance scores~$s_i$ for a set of passages $\{d_1, \dots d_k\}$ of which only one is a labeled relevant passage~$d_+$. The other $k-1$ passages are sampled from the top-200 passages retrieved by ColBERTv2~\cite{santhanam:2022} for~$q$.

In the second stage, the model is distilled from LLM-based cross-encoder rankings using the Rank-DistiLLM dataset~\cite{schlatt:2024} and the RankNet loss~\cite{burges:2005}:
\[
  \mathcal{L}_{\text{RankNet}} = \sum_{i=1}^{n} \sum_{j=1}^{n} \mathbbm{1}_{r_i < r_j} \log (1 + e^{s_{i} - s_{j}})\,,
\]
where~$\mathbbm{1}$ is the indicator function and $r_i$~is the rank the passage~$d_i$ gets assigned by a RankZephyr~\cite{pradeep:2023} teacher model, a state-of-the-art LLM-based cross-encoder.

\subsection{Fine-tuning to Detect Duplicates}
\label{sec:duplicate-aware-infonce}

In experiments on TREC Deep Learning and TIREx we found the Set-Encoder does not need passage interactions to be effective and, therefore, does not make use of the [INT]~token to share information between passages (details in Section~\ref{sec:cranfield-effectiveness}). To ensure that the model actually learns passage interactions, we modify the first-stage fine-tuning to include an inter-passage duplicate detection loss. We randomly sample a passage~$d_\times$ from the set of input passages, duplicate it, and add it to the set of input passages as passage $d_{k+1}$. We then add a binary classification head to the model that outputs a probability~$p_i$ of whether passage~$d_i$ is the duplicate passage. As a duplicate-aware~InfoNCE, we use InfoNCE plus the binary cross-entropy loss over the model's duplication probabilities:
\[
  \mathcal{L}_{\text{DA-InfoNCE}}  = \mathcal{L}_{\text{InfoNCE}} + \sum_{i=1}^{k} \mathbbm{1}_{d_i = d_\times} \log p_i + \mathbbm{1}_{d_i \neq d_\times} \log (1 - p_i)\,.
\]

\subsection{Fine-tuning to Rank According to Novelty}
\label{sec:novelty-aware-ranknet}

Finally, to demonstrate advantages of listwise models, we also fine-tune the Set-Encoder to re-rank based on passage relevance and novelty (i.e., whether a passage provides novel information compared to above ranked ones~\cite{clarke:2008}). As there are no large-scale datasets for this task (e.g., the diversity portion of the TREC~Web tracks is too small for fine-tuning), we create the new Rank-DistiLLM-Novelty dataset exploiting that MS~MARCO, the basis for Rank-DistiLLM, contains many near-duplicate passages. In a ranking from the Rank-DistiLLM dataset, we group passages into a ``novelty group'' that scikit-learn's~\cite{pedregosa:2011} agglomerative clustering combines with a word-based Jaccard similarity~$>0.5$. We treat the TREC~2019 and~2020 Deep Learning data analogously for evaluation~\cite{craswell:2019,craswell:2020}.

For novelty-aware re-ranking, we propose a new novelty-aware RankNet loss function. Intuitively, the loss penalizes a model if it ranks a less relevant passage higher than a more relevant one or a near-duplicate passage higher than a non-duplicate one. Let $c_i$ be the cluster of near-duplicate passages to which $d_i$ belongs. The novelty-aware RankNet loss then is
\[
  \mathcal{L}_{\text{NA-RankNet}} = \sum_{i=1}^{k} \sum_{j=1}^{k} \mathbbm{1}_{\bar{r}_i < \bar{r}_j} \log (1 + e^{s_i - s_j}) \,,
\]
with  $\bar{r}_i = \,\, r_i \cdot (1 - \max_{j=1 \dots k} \mathbbm{1}_{c_i = c_j} \cdot \mathbbm{1}_{s_i < s_j})$ denoting adjusted relevance labels that are set to zero if the model already has ranked a near-duplicate higher.
\section{Evaluation}
\label{evaluation}

To evaluate the effectiveness of the Set-Encoder, we conduct experiments on the TREC Deep Learning (DL) 2019 and 2020 passage tracks~\cite{craswell:2019,craswell:2020} and on all tasks from the 13~collections in the TIREx platform~\cite{frobe:2023,hashemi:2020,clarke:2009,clarke:2010,clarke:2011,clarke:2012,collins-thompson:2013,collins-thompson:2014,bondarenko:2021,bondarenko:2022,voorhees:2020,wang:2020,cleverdon:1991,voorhees:1998,voorhees:1999,voorhees:2004,craswell:2002,craswell:2003,craswell:2004,clarke:2004,clarke:2005,buttcher:2006,hersh:2004,hersh:2005,roberts:2017,roberts:2018,craswell:2019,craswell:2020,boteva:2016}. For TREC DL, we report nDCG@10 when re-ranking the top-$100$~passages retrieved by either BM25~\cite{robertson:1994} or ColBERTv2~\cite{santhanam:2022}. For the novelty-based ranking, we also report $\alpha$-nDCG@10~\cite{clarke:2008} and consider all passages from a near-duplicate cluster to belong to a subtopic (cf.~Section~\ref{sec:novelty-aware-ranknet} for the clustering). We set $\alpha = 0.99$, so that only the first passage from a subtopic contributes to the gain. For TIREx, we report nDCG@10 micro-averaged across all queries associated with a collection and we report the macro-averaged arithmetic and geometric mean across all collections.

We compare the Set-Encoder with RankGPT-4o~\cite{sun:2023} and RankZephyr~\cite{pradeep:2023}, two state-of-the-art listwise LLM-based cross-encoders. As GPT-4o has a maximum input length of $128,000$~tokens, we also test re-ranking $100$~passages at once~(RankGPT-4o~Full). Additionally, we include LiT5-Distill~\cite{tamber:2023}, mono\-T5~3B~\cite{nogueira:2020b}, and RankT5~3B~\cite{zhuang:2022} cross-encoders, and a pointwise monoELECTRA model as it basically is the Set-Encoder without inter-passage attention~\cite{schlatt:2024}.

\subsection{Experimental Setup}

We mostly follow \citet{schlatt:2024} for fine-tuning the Set-Encoder and monoELECTRA models. We truncate queries to at most $32$~tokens and passages to $256$~tokens and use ELECTRA\textsubscript{BASE}%
\footnote{\texttt{google/electra-base-discriminator}}
and ELECTRA\textsubscript{LARGE}%
\footnote{\texttt{google/electra-large-discriminator}}
\cite{clark:2020} checkpoints from HuggingFace~\cite{wolf:2020} as starting points for fine-tuning. We use the relevance labels from the MS~MARCO passage dataset~\cite{nguyen:2016} and the ColBERTv2~\cite{santhanam:2022} hard negatives from Rank-DistiLLM~\cite{schlatt:2024} for first-stage fine-tuning. We fine-tune all models for $20$k~steps with a batch size of~$32$ using InfoNCE loss and $7$~negatives  per query. We then use the ColBERTv2--RankZephyr distillation scores from RankDistiLLM for second-stage fine-tuning with a batch size of~$32$ and $100$~passages per query and fine-tune for $3$~epochs using RankNet loss. For both fine-tuning stages we use the AdamW~\cite{loshchilov:2019} optimizer with a learning rate of $10^{-5}$.

We also fine-tune the Set-Encoder in the first-stage using our newly proposed duplicate-aware InfoNCE for at least $20$k~steps but at most $60$k steps or until the duplicate-detection cross-entropy loss is smaller than $0.05$ for at least $100$~steps. For the novelty ranking task, we fine-tune models for $10$~epochs using our novelty-aware RankNet loss, using early stopping and $\alpha$-nDCG@10 on the TREC Deep Learning 2021 task as the criterion.

All models were fine-tuned on a single NVIDIA A100 40GB GPU and we used the following packages and frameworks: ir\_datasets, ir-measures, Jupyter, Lightning, Lightning~IR, matplotlib, NumPy, pandas, PyTerrier, PyTorch, and SciPy~\cite{kluyver:2016, falcon:2023, schlatt:2024c, hunter:2007, harris:2020, pandas:2024, macdonald:2021, paszke:2019, virtanen:2020, macavaney:2021, macavaney:2022}.

\begin{table}[t]
  \centering
  \scriptsize
  \setlength{\tabcolsep}{7pt}
  \caption{Effectiveness~(nDCG@10) of various cross-encoders on the TREC~Deep Learning~2019 and 2020 tracks using BM25 or ColBERTv2 as the first stage retrieval. The highest and second-highest scores per setting are bold and underlined, respectively. Model sizes given as number of parameters. \sig~denotes a significant difference ($p < 0.05$, Holm-Bonferroni-corrected) to the Set-Encoder with 330M parameters.}
  \begin{tabular}{@{}lrcccc@{}}
    \toprule
    \bfseries Model              & \bfseries Size & \multicolumn{2}{c}{\bfseries TREC DL 19} & \multicolumn{2}{c}{\bfseries TREC DL 20}                                         \\
    \cmidrule(l@{\tabcolsep}r@{\tabcolsep}){3-4} \cmidrule(l@{\tabcolsep}){5-6}
                                 &                & BM25                                     & CBv2                                     & BM25              & CBv2              \\
    \midrule
    First Stage                  & --             & 0.480\kernSig                            & 0.732\kernSig                            & 0.494\kernSig     & 0.724\kernSig     \\
    \midrule
    RankGPT-4o                   & N/A            & 0.725                                    & 0.784                                    & 0.719             & 0.793             \\
    RankGPT-4o Full              & N/A            & \underline{0.732}                        & 0.781                                    & 0.711             & 0.796             \\
    RankZephyr                   & 7B             & 0.719                                    & 0.749                                    & 0.720             & \underline{0.798} \\
    LiT5-Distill                 & 220M           & 0.696                                    & 0.753                                    & 0.679\kernSig     & 0.744\kernSig     \\
    monoT5 3B                    & 3B             & 0.705                                    & 0.745                                    & 0.715             & 0.757             \\
    RankT5 3B                    & 3B             & 0.710                                    & 0.752                                    & 0.711             & 0.772             \\
    \midrule
    \multirow{2}{*}{monoELECTRA} & 110M           & 0.720                                    & 0.768                                    & 0.711\kernSig     & 0.770             \\
                                 & 330M           & \textbf{0.733}                           & 0.765                                    & \underline{0.727} & \textbf{0.799}    \\
    \midrule
    \multirow{2}{*}{Set-Encoder} & 110M           & 0.724                                    & \underline{0.788}                        & 0.710\kernSig     & 0.777             \\
                                 & 330M           & 0.727                                    & \textbf{0.789}                           & \textbf{0.735}    & 0.790             \\
    \bottomrule
  \end{tabular}
  \label{tbl:trec-dl-results}
\end{table}

\subsection{Effectiveness on Cranfield-Style Datasets}
\label{sec:cranfield-effectiveness}

\paragraph{In-domain Re-ranking.}
\label{par:in-domain-re-ranking}
Table~\ref{tbl:trec-dl-results} shows the models' effectiveness on TREC~DL. The Set-Encoder is on par with the state-of-the-art cross-encoders---the 330M parameter variant even is the most effective in two of four retrieval settings and the differences to the other cross-encoders is not significant in any of the scenarios (paired t-test with $p < 0.05$ and Holm-Bonferroni correction). Interestingly, the pointwise monoELECTRA model is similarly effective. This suggests that passage interactions are unnecessary; a stark contrast to previous work stating that passage interactions are beneficial for re-ranking~\cite{nogueira:2019,nogueira:2020b,tamber:2023,pradeep:2021,zhuang:2022,pradeep:2023}.

\begin{table}[t]
  \centering
  \scriptsize
  \renewcommand{\tabcolsep}{1.83pt}
  \newcolumntype{R}[3]{%
    >{\adjustbox{right=#1,angle=#2,lap=#3-\width}\bgroup}%
    c%
    <{\egroup}%
  }
  \newcommand{\rot}[4]{\multicolumn{1}{R{#1}{#2}{#3}}{#4}}
  \caption[]{Effectiveness~(nDCG@10) of various cross-encoders micro-averaged for test collections from the TIREx framework~\cite{frobe:2023}. The macro-averaged geometric means (column `G. Mean') are computed across all collections. Model sizes given as the number of parameters. The highest and second-highest averaged scores per collection are bold and underlined, respectively. \sig~denotes a significant difference ($p < 0.05$, Holm-Bonferroni-corrected) to the Set-Encoder with 330M parameters.\protect\footnotemark}
  \begin{tabular}{@{}lrcccccccccccccc@{}}
    \toprule
    \raisebox{-7ex}{\bfseries Model}
                                 & \raisebox{-7ex}{\kern-2em\bfseries Size}
                                 & \rot{5.2em}{-30}{1.5em}{\bfseries Antique}
                                 & \rot{5.2em}{-30}{1.5em}{\bfseries Args.me}
                                 & \rot{5.2em}{-30}{1.5em}{\bfseries ClueWeb09}
                                 & \rot{5.2em}{-30}{1.5em}{\bfseries ClueWeb12}
                                 & \rot{5.2em}{-30}{1.5em}{\bfseries CORD-19}
                                 & \rot{5.2em}{-30}{1.5em}{\bfseries Cranfield}
                                 & \rot{5.2em}{-30}{1.5em}{\bfseries Disks4+5}
                                 & \rot{5.2em}{-30}{1.5em}{\bfseries GOV}
                                 & \rot{5.2em}{-30}{1.5em}{\bfseries GOV2}
                                 & \rot{5.2em}{-30}{1.5em}{\bfseries MEDLINE}
                                 & \rot{5.2em}{-30}{1.5em}{\bfseries NFCorpus}
                                 & \rot{5.2em}{-30}{1.5em}{\bfseries Vaswani}
                                 & \rot{5.2em}{-30}{1.5em}{\bfseries WaPo}
                                 & \rot{5.2em}{-30}{1.5em}{\bfseries G.\ Mean}                                                                                                                                                                                                                                                                                           \\
    \midrule
    First Stage                  & --                                           & .516\kernSig     & .405             & .177\kernSig     & \textbf{.364}\kernSig & .586\kernSig          & \textbf{.012}    & .424\kernSig     & .259\kernSig     & .467\kernSig     & .385                  & .281\kernSig     & .447\kernSig     & .364\kernSig     & .286             \\
    \midrule
    RankZephyr                   & 7B                                           & .534\kernSig     & .364\kernSig     & .213             & .303                  & \textbf{.767}\kernSig & .009             & \underline{.542} & \underline{.349} & .560             & \textbf{.460}\kernSig & .314             & .512             & \textbf{.508}    & .320             \\
    LiT5-Distill                 & 220M                                         & .576\kernSig     & .395             & .214             & .275\kernSig          & .686                  & \underline{.011} & .495\kernSig     & .304\kernSig     & .534\kernSig     & .354\kernSig          & .293\kernSig     & .429\kernSig     & .470             & .302             \\
    monoT5 3B                    & 3B                                           & .590             & \underline{.415} & .188\kernSig     & .323                  & .649\kernSig          & \underline{.011} & .526             & .345             & .529\kernSig     & .395                  & \underline{.319} & .474\kernSig     & .469             & .313             \\
    RankT5 3B                    & 3B                                           & \underline{.598} & \textbf{.421}    & \textbf{.227}    & \underline{.336}      & .713                  & .010             & .538             & \textbf{.353}    & .528\kernSig     & .406                  & \textbf{.323}    & .459\kernSig     & .468\kernSig     & \textbf{.322}    \\
    \midrule
    \multirow{2}{*}{m.ELECTRA}   & 110M                                         & .593             & .375\kernSig     & .209             & .295                  & .692                  & .010             & .507\kernSig     & .305\kernSig     & .541\kernSig     & .399                  & .306             & .522             & .458\kernSig     & .309             \\
                                 & 330M                                         & .575\kernSig     & .369\kernSig     & .221             & .313                  & \underline{.716}      & .008             & \textbf{.546}    & .344             & \underline{.572} & \underline{.419}      & .316             & \underline{.526} & \underline{.504} & .318             \\
    \midrule
    \multirow{2}{*}{Set-Encoder} & 110M                                         & .594             & .375\kernSig     & .216             & .299                  & .683                  & .010             & .513\kernSig     & .306\kernSig     & .543\kernSig     & .396                  & .306             & .523             & .461\kernSig     & .311             \\
                                 & 330M                                         & \textbf{.606}    & .409             & \underline{.226} & .310                  & .702                  & .009             & .534             & .334             & \textbf{.573}    & .405                  & .313             & \textbf{.530}    & \textbf{.508}    & \underline{.321} \\
    \bottomrule
  \end{tabular}
  \label{tbl:tirex-results}
\end{table}

\paragraph{Out-of-domain Re-ranking.}
\label{par:out-of-domain-re-ranking}
The in-domain findings also hold in out-of-domain scenarios. Table~\ref{tbl:tirex-results} shows the effectiveness on the 13~TIREx collections (RankGPT excluded as TIREx does not allow external API~calls) and the Set-Encoder is again on par with the state-of-the-art RankZephyr and RankT5. The differences are significant in some cases, but on average the models are similarly effective.

\paragraph{Passage Interaction ``Unneccessity''.}
The on-par monoELECTRA effectiveness in the above experiments suggests that passage interactions are not really necessary. We think that there are three main reasons. First, we have fine-tuned monoELECTRA with the best available data~\cite{schlatt:2024}, second, the TREC~DL and TIREx scenarios come with relevance judgments that are independent of each other, and third, the standard evaluation schemes do not discount near-duplicate results. As the assessors and evaluation treat passages independent of each other, re-rankers very likely can also score passages individually and still be effective. Passage interactions are beneficial in tasks where, for example, fairness, diversity, or novelty are considered, and we investigate this in more detail in Section~\ref{sec:novelty-ranking}.

\paragraph{Permutation Invariance.}
Additionally, despite being fine-tuned on RankZephyr rankings, the Set-Encoder and monoELECTRA are more effective in most out-of-domain re-ranking tasks. The Set-Encoder achieves a higher effectiveness than RankZephyr on $6$~out of $13$~corpora, RankZephyr is more effective on $5$~corpora, and they reach same effectiveness on $2$. Also monoELECTRA is more effective than RankZephyr on $9$~corpora and less effective on only~$4$. RankZephyr only reaches a comparable mean effectiveness by being substantially more effective on the two medical corpora, CORD-19 and MEDLINE. We attribute the teacher model, RankZephyr, to be less effective than the student models to the fact that it is not permutation invariant. Consequently, RankZephyr is sensitive to the quality of the first-stage retrieval, which we investigate further in Section~\ref{sec:permutation-invariance}.

\footnotetext{This table very slightly differs from the table in the \href{https://link.springer.com/chapter/10.1007/978-3-031-88711-6_1}{official proceedings}: it correctly reports micro-averaged results and the geometric mean.}

\subsection{Novelty-Aware Ranking}
\label{sec:novelty-ranking}

\begin{table}[t]
  \centering
  \scriptsize
  \setlength{\tabcolsep}{5pt}
  \caption{Effectiveness (nDCG@10 and $\alpha$-nDCG@10, $\alpha=0.99$) of cross-encoders on the TREC~Deep Learning~2019 and 2020 tracks clustered into near-duplicate subtopics. The highest and second-highest scores per setting are bold and underlined, respectively. \sig~denotes a significant difference ($p<0.05$, Holm-Bonferroni-corrected) to the Set-Encoder fine-tuned using $\mathcal{L}_{\text{DA-InfoNCE}}$ and then $\mathcal{L}_{\text{NA-RankNet}}$ (Row 14).}
  \begin{tabular}{@{}llllcccc@{\hspace{\widthof{\sig}}}}
    \toprule
         & \bfseries Model                    & \multicolumn{2}{l}{\bfseries Prompt / Loss}        & \multicolumn{2}{c}{\bfseries nDCG} & \multicolumn{2}{c}{\bfseries $\alpha$-nDCG}                                                      \\
    \cmidrule(l@{\tabcolsep}r@{\tabcolsep}){5-6} \cmidrule(l@{\tabcolsep}){7-8}
         &                                    &                                                    &                                    & 2019                                        & 2020          & 2019              & 2020           \\
    \midrule
    (1)  & First Stage                        & \multicolumn{2}{l}{--}                             & 0.732                              & 0.724                                       & 0.700\kernSig & 0.722\kernSig                      \\
    \midrule
    (2)  & \multirow{2}{*}{RankGPT-4o}        & \multicolumn{2}{l}{Relevance}                      & 0.784\kernSig                      & 0.793\kernSig                               & 0.750         & 0.759                              \\
    (3)  &                                    & \multicolumn{2}{l}{Novelty}                        & 0.778\kernSig                      & \textbf{0.806}\kernSig                      & 0.741         & \underline{0.773}                  \\
    \midrule
    (4)  & \multirow{2}{*}{RankGPT-4o Full}   & \multicolumn{2}{l}{Relevance}                      & 0.781\kernSig                      & 0.796\kernSig                               & 0.738         & 0.763                              \\
    (5)  &                                    & \multicolumn{2}{l}{Novelty}                        & \underline{0.785}\kernSig          & \underline{0.803}\kernSig                   & 0.750         & 0.771                              \\
    \midrule
    (6)  & \multirow{2}{*}{RankZephyr}        & \multicolumn{2}{l}{Relevance}                      & 0.749                              & 0.798\kernSig                               & 0.699\kernSig & 0.765                              \\
    (7)  &                                    & \multicolumn{2}{l}{Novelty}                        & 0.753                              & 0.800\kernSig                               & 0.700\kernSig & 0.760                              \\
    \midrule
    (8)  & \bfseries Model                    & \bfseries 1st $\mathbfcal{L}$                      & \bfseries 2nd $\mathbfcal{L}$      &                                             &                                                    \\
    \midrule
    (9)  & \multirow{2}{*}{monoELECTRA}       & \multirow{2}{*}{$\mathcal{L}_{\text{InfoNCE}}$}    & $\mathcal{L}_{\text{RankNet}}$     & 0.768\kernSig                               & 0.770\kernSig & 0.718\kernSig     & 0.745\kernSig  \\
    (10) &                                    &                                                    & $\mathcal{L}_{\text{NA-RankNet}}$  & 0.704                                       & 0.675         & \underline{0.785} & 0.753          \\
    \midrule
    (11) & \multirow{4}{*}[-1ex]{Set-Encoder} & \multirow{2}{*}{$\mathcal{L}_{\text{InfoNCE}}$}    & $\mathcal{L}_{\text{RankNet}}$     & 0.780\kernSig                               & 0.757\kernSig & 0.733\kernSig     & 0.747\kernSig  \\
    (12) &                                    &                                                    & $\mathcal{L}_{\text{NA-RankNet}}$  & 0.714                                       & 0.651         & 0.779             & 0.743\kernSig  \\
    \cmidrule(l@{\tabcolsep}r@{\tabcolsep}){3-8}
    (13) &                                    & \multirow{2}{*}{$\mathcal{L}_{\text{DA-InfoNCE}}$} & $\mathcal{L}_{\text{RankNet}}$     & \textbf{0.788}\kernSig                      & 0.777\kernSig & 0.740\kernSig     & 0.752\kernSig  \\
    (14) &                                    &                                                    & $\mathcal{L}_{\text{NA-RankNet}}$  & 0.710                                       & 0.690         & \textbf{0.821}    & \textbf{0.803} \\
    \midrule
    (15) & Set-Enc. \st{[INT]}                & $\mathcal{L}_{\text{DA-InfoNCE}}$                  & $\mathcal{L}_{\text{NA-RankNet}}$  & 0.707                                       & 0.670         & 0.773             & 0.748\kernSig  \\
    \bottomrule
  \end{tabular}
  \label{tbl:trec-dl-novelty-results}
\end{table}

To investigate the potential of passage interactions, we task the models to also take novelty into account (i.e., from a group of relevant but near-duplicate passages, only one should be ranked high). Using our novelty-aware RankNet loss (see Section~\ref{sec:novelty-aware-ranknet}), we fine-tune the Set-Encoder and monoELECTRA on novelty and we fine-tune the Set-Encoder on duplicate detection using our duplicate-aware InfoNCE~loss (cf.~Section~\ref{sec:duplicate-aware-infonce}; monoELECTRA is incapable of detecting duplicates). For the LLM-based cross-encoders, we augment the prompt to account for novelty. Table~\ref{tbl:trec-dl-novelty-results} shows the nDCG@10 and $\alpha$-nDCG@10 of the models.

The LLM-based cross-encoders are largely unaffected by the novelty prompt. Only RankGPT-4o Full has marginally but consistently higher $\alpha$-nDCG@10 using the novelty prompt compared to the relevance prompt. One reason probably is the windowed ranking strategy of LLM-based cross-encoders. As the windowed strategy starts from the bottom of a ranking, initially highly ranked passages that are near-duplicates of other highly ranked passages simply cannot be moved to low ranks (cf.~Section~\ref{sec:permutation-invariance} for a more detailed analysis).

In contrast, the Set-Encoder improves considerably when fine-tuned using the novelty-aware NA-RankNet. It is significantly more effective in terms of $\alpha$-nDCG@10 than its counterpart (compare rows 13 and 14) fine-tuned using normal RankNet. While not all $\alpha$-nDCG@10 differences to the other cross-encoders are significant, the differences to the second-best models are sizable and are the largest between any two neighboring models when sorted by $\alpha$-nDCG@10.

Looking at the InfoNCE-based fine-tuning, we find that the Set-Encoder fine-tuned with standard~InfoNCE (row 12) is on par with a pointwise monoELECTRA model (row 10) in novelty-aware ranking. That is, the Set-Encoder is unable to profit from inter-passage attention when first fine-tuned to only rank according to relevance. However, when initially fine-tuned using duplicate-aware DA-InfoNCE (i.e., when the model learns to rank according to relevance and to detect exact duplicates; row 14), the Set-Encoder profits from inter-passage attention and is substantially more effective. We interpret this as further evidence that typical TREC-style relevance assessments and evaluation setups do not contain a training signal which necessitates passage interactions. Therefore, trained on such data, the Set-Encoder does not learn to use inter-passage attention. Only when the model is specifically ``primed'' to use inter-passage attention, by being fine-tuned to detect duplicates, can it learn passage interactions and then improve on novelty ranking.

\paragraph{[INT] Ablation.} To determine the effect of the [INT]~token, we compare the Set-Encoder's effectiveness when using the [INT]~token and when using the [CLS]~token for interaction. The results in the row 15 of Table~\ref{tbl:trec-dl-novelty-results} indicate that the model using the [CLS]~token does not model passage interactions since it is on par with the pointwise monoELECTRA (row 10). In fact, using the [CLS]~token instead of the [INT]~token, the model is already unable to learn to detect duplicates during the initial fine-tuning. We hypothesize that the [CLS]~token is not suited to model passage interactions as it has been heavily pre-trained and thus is not flexible enough to learn new interactions in fine-tuning. Further investigating this hypothesis could be an interesting direction for future work.

\subsection{Permutation Invariance}
\label{sec:permutation-invariance}

\begin{figure}[t]
\centering
\includegraphics[width=\linewidth]{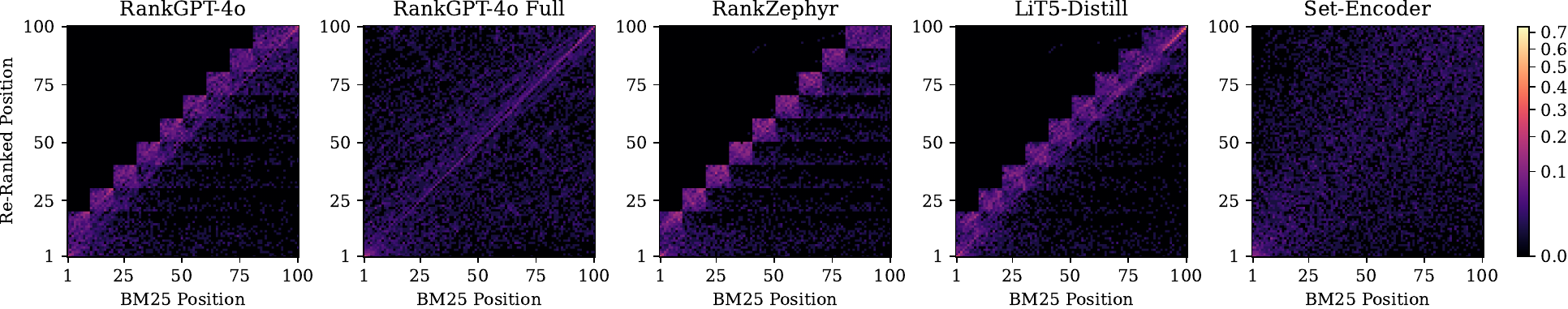}
\caption{Proportional rank changes of various cross-encoders for re-ranking BM25 averaged across all queries from the TREC Deep Learning 2019 and 2020 tracks.}
\label{fig:rank-changes}
\end{figure}

\paragraph{Positional Biases.} All previous listwise cross-encoders have an implicit positional bias by concatenating the input passages. Most LLM-based cross-encoders also have an explicit positional bias due to the limited input length. For example, to re-rank $100$~passages, the original RankGPT model, RankZephyr, and LiT5-Distill use an overlapping windowed strategy. Starting from the bottom of the ranking, the models re-rank the passages from positions $100$~to~$81$, then from $90$~to~$71$, and so on until the top $20$~passages are re-ranked. We visualize these explicit and implicit biases in Figure~\ref{fig:rank-changes} as the average proportional rank changes when re-ranking the top-100 passages retrieved by BM25 for TREC~DL~2019 and~2020. The lighter a pixel, the more frequently a model ranks a passage from a particular position to another particular position.

For RankGPT-4o, RankZephyr, and LiT5-Distill, the explicit bias from the windowed strategy is immediately visible as a distinct step pattern. RankGPT-4o Full, which does not use the windowed strategy and re-ranks all 100~passages at once, does not exhibit the step pattern but instead has a distinct diagonal line that indicates that the model tends to keep passages in the same position as in the original ranking, caused by the model's auto-regressiveness. It is tasked to output the order of passages in the form of bracketed positions (e.g., \texttt{[2] > [1] > [3] > \ldots}) and has an implicit bias to simply output sequential positions.

\begin{figure}[t]
\begin{minipage}[t]{0.48\textwidth}
  \vspace{0pt}
  \centering
  \captionof{figure}{Effectiveness~(nDCG@10) on different permutations of BM25 rankings for TREC Deep Learning 2019 and 2020.}
  \label{fig:permutation-effectiveness}
  \includegraphics[width=\textwidth]{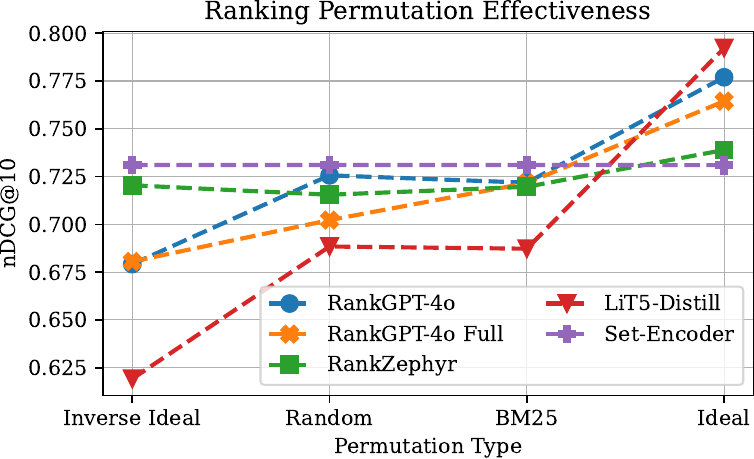}
\end{minipage}
\hfill
\begin{minipage}[t]{0.48\textwidth}
  \vspace{0pt}
  \scriptsize
\centering
\setlength{\tabcolsep}{3.2pt}
\captionof{table}{Time~(sec.) and memory footprint~(GB) for re-ranking 100~passages.}
\begin{tabular}{@{}lrrr@{}}
  \toprule
  \bfseries Model              & \bfseries Size & \bfseries Time & \bfseries Memory \\
  \midrule
  RankGPT-4o                   & N/A            & 18.773         & N/A              \\
  RankGPT-4o Full              & N/A            & 7.362          & N/A              \\
  RankZephyr                   & 7B             & 24.047         & 15.48            \\
  LiT5-Distill                 & 220M           & 2.054          & 2.69             \\
  monoT5\textsubscript{3B}     & 3B             & 0.998          & 29.36            \\
  RankT5\textsubscript{3B}     & 3B             & 0.942          & 29.04            \\
  \midrule
  \multirow{2}{*}{monoELECTRA} & 110M           & 0.139          & 1.18             \\
                               & 330M           & 0.215          & 2.69             \\
  \midrule
  \multirow{2}{*}{Set-Encoder} & 110M           & 0.147          & 1.25             \\
                               & 330M           & 0.219          & 2.60             \\
  \bottomrule
\end{tabular}
\label{tab:efficiency}
\end{minipage}
\end{figure}

By contrast, the Set-Encoder does not exhibit any immediately obvious positional biases; only a slight clustering of lighter pixels in the lower left corner and darker areas in the upper left and lower right show that the Set-Encoder's top-results correlate to some degree with the ones from the original~BM25.

\paragraph{Ranking Perturbations.} To test how the positional biases affect a cross-encoder's ranking effectiveness, we generate several ``corrupted'' perturbations of the top-$100$ passages retrieved by BM25 on the TREC DL 2019 and 2020 tracks: a random permutation, an ideal permutation, and a reverse-ideal permutation. Ideal and reverse-ideal permutations are generated by reordering the passages according to their relevance judgments. Figure~\ref{fig:permutation-effectiveness} visualizes the nDCG@10 of various LLM-based cross-encoders and our Set-Encoder on the different permutations.

The Set-Encoder is robust to the order of input passages and has the same effectiveness irrespective of the permutation. All other listwise cross-encoders are affected by the ranking permutations. RankGPT-4o, RankGPT-4o Full, and LiT5-Distill lose a substantial portion of their effectiveness when re-ranking the inverse ideal ranking and improve with increasingly better initial rankings. Only RankZephyr is comparatively robust to perturbations because it is fine-tuned to counteract positional biases, but it nonetheless fluctuates slightly.

Overall, the Set-Encoder reaches a higher nDCG@10 than the other cross-encoders on the inverse ideal, on the random, and on the original BM25~rankings. The LLM-based cross-encoders only achieve higher effectiveness when re-ranking the already ideal permutation. Due to the positional biases, most LLM-based cross-encoders require a good initial ranking to be effective. This insight explains how the Set-Encoder and monoELECTRA can manage to be more effective than their RankZephyr teacher model when re-ranking BM25~rankings for both the in-domain TREC Deep Learning scenarios and for most collections in the out-of-domain TIREx scenarios from Section~\ref{sec:cranfield-effectiveness}.

\section{Efficiency Analysis}
\label{sec:efficiency-analysis}

As for the models' efficiency, we report inference time and GPU memory footprint when re-ranking TREC DL. The results in Table~\ref{tab:efficiency} show that the Set-Encoder adds only minimal overhead over monoELECTRA (i.e., over a similar model but without inter-passage attention). The Set-Encoder models are slightly slower than the respective monoELECTRA counterparts and the 110M~parameter model uses marginally more memory. Interestingly, the 330M~Set-Encoder uses slightly less memory than monoELECTRA, despite adding the additional [INT]~interaction tokens. As this result is consistent across different batch sizes, we hypothesize that the CUDA-backend may be able to use more efficient kernels, but leave deeper investigations of the effect to future work.

Compared to state-of-the-art listwise LLM-based cross-encoders, the Set-Encoder substantially improves efficiency. While achieving comparable or better effectiveness  (Sections~\ref{sec:cranfield-effectiveness} and~\ref{sec:permutation-invariance}), the 330M~Set-Encoder is around 85~times faster than RankGPT-4o using a windowed strategy, about 33~times faster than RankGPT-4o re-ranking 100~passages at once, and about 110~times faster than RankZephyr. RankGPT-4o's memory consumption is not available, but compared to RankZephyr, the Set-Encoder uses 6~times less memory. Comparing with the smaller LLM-based LiT5-Distill cross-encoder, the Set-Encoder has the same memory footprint but is 9.5~times faster and substantially more effective.
\section{Conclusion}

In this paper, we have introduced the Set-Encoder: a new cross-encoder architecture that models passage interactions in a permutation-invariant way. Contrasting previous work, our empirical results show that passage interactions do not necessarily improve a cross-encoder's re-ranking effectiveness. A pointwise model fine-tuned using current distillation methods is as effective as state-of-the-art LLM-based cross-encoders in typical Cranfield-style settings in which the passages were judged independent of each other so that passage interactions are not important for effective re-ranking. But when additionally considering inter-document related aspects such as novelty, the Set-Encoder is more effective than its pointwise counterpart and previous state-of-the-art listwise cross-encoders.

Our results also show that permutation invariance is very helpful for an efficient and effective re-ranking. Explicit and implicit positional biases in previous listwise cross-encoders lead to inconsistent and suboptimal effectiveness across different settings. Using our new permutation-invariant inter-passage attention pattern, the Set-Encoder is on par or more effective than previous listwise cross-encoders in all ranking settings. At the same time, the Set-Encoder is way more efficient in terms of both memory consumption and query latency.

{\fontsize{9pt}{11pt}\selectfont 
\subsubsection*{Acknowledgements}
This publication has received funding from the European Union's Horizon Europe research and innovation programme under grant agreement No 101070014 (OpenWebSearch.EU, \url{https://doi.org/10.3030/101070014}).
}

\begin{raggedright}
  \small
  \bibliography{ecir25-set-encoder-lit}
\end{raggedright}

\end{document}